# Observation of Pull-in Instability in Graphene Membranes under Interfacial Forces

*Xinghui Liu[1], Narasimha G. Boddeti[1], Mariah R. Szpunar[2], Luda Wang[1], Miguel A. Rodriguez[3], Rong Long[1, 4], Jianliang Xiao[1], Martin L. Dunn[5], and J. Scott Bunch[1]\**

[1]Department of Mechanical Engineering, University of Colorado, Boulder, CO 80309 USA

[2]Department of Mechanical Engineering, University of Miami, Coral Gables, FL 33124 USA

[3]Department of Mechanical Engineering, Columbia University, New York, NY 10027

[4]Department of Mechanical Engineering, University of Alberta, Edmonton, Alberta T6G 2G8, Canada

[5]Singapore University of Technology and Design, Singapore, 138682

*email: jbunch@colorado.edu

Abstract

We present a unique experimental configuration that allows us to determine the interfacial forces on nearly parallel plates made from the thinnest possible mechanical structures, single and few layer graphene membranes. Our approach consists of using a pressure difference across a graphene membrane to bring the membrane to within ~ 10-20 nm above a circular post covered with $SiO_x$ or Au until a critical point is reached whereby the membrane snaps into adhesive contact with the post. Continuous measurements of the deforming membrane with an AFM coupled with a theoretical model allow us to deduce the magnitude of the interfacial forces between graphene and $SiO_x$ and graphene and Au. The nature of the interfacial forces at ~ 10 - 20 nm separations is consistent with an inverse fourth power distance dependence, implying that the interfacial forces are dominated by van der Waals interactions. Furthermore, the strength of the interactions is found to increase linearly with the number of graphene layers. The experimental approach can be used to measure the strength of the interfacial forces for other atomically thin two-dimensional materials, and help guide the development of nanomechanical devices such as switches, resonators, and sensors.

Interfacial forces act between all materials[1]. At macroscopic distances, these interfacial forces are weak and practically insignificant, but at distances approaching tens of nanometers, they become much stronger, thereby enhancing the attraction within micro/nanomechanical structures or molecules, and potentially significantly affecting the device performance[2-5]. Graphene, a 2 dimensional nanomaterial composed of carbon atoms, is a promising material with potential applications in a variety of nanomechanical, biological and electrical devices due to its exceptional properties[6-14]. Furthermore, graphene being extremely thin with a very high surface area to volume ratio is highly susceptible to interfacial forces and is an ideal candidate to study and characterize these forces[15, 16]. Therefore, there is an increasing interest in studying the nature of interfacial forces on graphene[17]. Even though the adhesion strength between graphene and substrates when in contact has been experimentally measured in different ways, experimental measurements of non-contact attractive interfacial forces remains relatively unexplored[18-21]. Interfacial forces on bulk materials or other nanomaterials have been measured using a variety of configurations[1, 4, 5, 22]. Here, we demonstrate a novel experimental method to study these elusive forces on graphene with a real time observation of the induced pull in instability.

Devices used in this study consist of a graphene flake suspended over an annular ring etched into a silicon oxide wafer, forming a graphene-sealed microcavity (Fig. 1a). Device configurations include graphene suspended on bare $SiO_x$ or gold-coated $SiO_x$. The graphene membranes are pressurized using a previously-developed technique[7, 18]. The suspended graphene membranes are placed in a high pressure chamber at a charging pressure, $p_{ext}$ ~ 300 kPa of $H_2$ gas, and left for a sufficiently long time (~10 hours) until the pressures inside, $p_{int}$, and outside of the microcavity, $p_{ext}$, equilibrate. After removing the sample from the high pressure chamber and bringing it to atmospheric pressure, a pressure difference, $\Delta p = p_{int} - p_{ext}$, exists across the graphene membrane. At low $\Delta p$, the graphene sheet remains adhered to the inner post and deforms in a donut shape (Fig. 1b). At sufficiently high $\Delta p$, the force is large enough to overcome the adhesion energy of the graphene to the inner post, and the graphene membrane delaminates from it, becoming a spherical cap (Fig. 1c).

After creating deformed spherical caps, our strategy is to then let gas slowly diffuse out of the microcavity through the underlying $SiO_x$ substrate which decreases $\Delta p$ and the corresponding central deflection, $h$, of the graphene membrane until it is pulled back onto the center post due to attractive interactions between the post and graphene membrane. This process is monitored in real-time using an atomic force microscope, AFM (Fig. 1d and supplementary movie). Figure 1d shows a series of AFM line scans through the center of a pressurized graphene membrane before and after the pull-in process. Initially a line trace through the center of the membrane (dark blue) corresponds to the situation in Fig. 1c where the graphene is delaminated from the inner post. At a later time (black) the graphene is pulled onto the post and the graphene is deformed in a donut shape as seen in Fig. 1b. The red line corresponds to a line trace just before pull-in. We call the center deflection at this point in time, the pull-in distance, $h_0$. Figure 1e shows the measured pull-in distance, $h_0$, vs. number of graphene layers for graphene sheets in an identical geometry on the same chip (see Supplementary Information). The number of graphene sheets was verified by Raman spectroscopy (see Supplementary Information). The pull-in distance measured on bare SiOx substrate, $h_0$, increases slightly with the # of layers from an average value of $h_0 = 9.2$ nm for 1 layer graphene to $h_0 = 10.8$ nm for 5 layer graphene. At these values of $h_0$, the variation in the height of the graphene over the post is small and the post and graphene are effectively 2 parallel plates.

The pull-in behavior observed here is similar to the pull-in or jump-in of a cantilever spring into contact due to interfacial forces[4, 23]. We model the pull-in behavior in a continuum setup by considering an isotropic pressurized graphene membrane with initial surface tension, $S_0$, and an attractive pressure, $P_{att}$, due to the interfacial force between the post and the graphene membrane[18, 24-27]. The analysis culminates in a relationship between the system parameters given by:

$$\frac{Et}{32\,a^2(1-v)}\left((\Delta p - P_{att})^2 b^4 + \Delta p^2(a^4 - b^4) + P_{att}^2 b^4 \log\left(\frac{a^4}{b^4}\right)\right.$$

$$\left. - 4\,\Delta p\, P_{att} b^2(a^2 - b^2)\right)$$

$$+ \left(S_0\left(\frac{1}{4h}\left(\Delta p\, a^2 - P_{att} b^2\left(1 + \log\left(\frac{a^2}{b^2}\right)\right)\right)\right)^2\right)$$

$$= \left(\frac{1}{4h}\left(\Delta p\, a^2 - P_{att} b^2\left(1 + \log\left(\frac{a^2}{b^2}\right)\right)\right)\right)^3 \quad (1)$$

where, $E$ is the elastic modulus of graphene, $t$ is the thickness, $v$ is the Poisson ratio, and $a$ and $b$ are the outer and inner radii of the annular cavity, respectively. Equation (1) establishes a relationship between $h$ and $\Delta p$ if $S_0$, $Et$, $a$, $b$, and $P_{att}$ are known. The radii, $a$ and $b$, are measured by AFM, while $Et$ and $v$ are taken from well-established values in the literature for single and few layer graphene[7, 18, 27]. We cannot directly measure $S_0$ so we assume values in the range of $S_0 = 0.03$ - $0.15$ N/m with an average value of $S_0 = 0.07$ N/m, consistent with numerous experimental measurements for exfoliated suspended graphene membranes in a similar geometry[7, 28, 29]. Figure 2a shows the relationship between $h$ vs. $\Delta p$ obtained from equation (1) using the system parameters for a monolayer graphene membrane: $a = 1.5$ µm, $b = 0.25$ µm, $S_0 = 0.07$ N/m, $Et = 340$ N-m, $v = 0.16$, and $P_{att} = \beta/h^4 = 0.0199$ nN-nm$^2$/h$^4$. The deflection, $h$ decreases with decreasing $\Delta p$ (leaking gas) until a critical point is reached. At this critical maximum deflection, $h_0$, the graphene is sufficiently close to the post and pulled into the post by the attractive force. This pull-in instability is illustrated by the point on the curve where the slope goes to infinity at the pull in distance $h_0$, or:

$$\frac{d\Delta p}{dh}\bigg|_{h=h_0} = 0 \qquad (2)$$

The measured $h_0$, $a$, and $b$, coupled with the values of $S_0$, $Et$, and $v$ taken from the literature, allow us to determine $P_{att}$ by solving eqs. (1) and (2) simultaneously for $\Delta p$ and $P_{att}$. A comparison to a high-fidelity finite element model that more accurately treats the spatial dependence of the attractive forces is shown in blue on Fig. 2a; the close agreement between them supports the validity of our analytical model.

We assumed an attractive force law of the form $P_{att} = \beta/h^4$, consistent with the van der Waals (vdW) force derived from Lifshitz theory between graphene and SiO$_2$ for separations on the order of 10 nm or the phenomenological Lennard-Jones pair potential of interaction [1, 16, 30-32]. From the experimentally measured pull-in distances in Fig. 1e we calculate $\beta$ for each device and arrive at the corresponding $P_{att}(h = h_0)$. This is shown in Fig. 2b where $\beta = 0.0199$ nN-nm$^2$ for monolayer graphene. This value is ~1.5% of the dispersion force between 2 perfectly metallic parallel plates, $P_{att} = \pi\hbar c/240h^4 = 1.3$ nN-nm$^2$/h$^4$ [33], and agrees reasonably well with recent theoretical calculations for graphene and SiO$_2$ at 10 nm separations, $\beta = 0.001$ nN-nm$^2$ - 0.01 nN-nm$^2$ for an intrinsic graphene doping density of $10^{14}$ m$^{-2}$ and $10^{16}$ m$^{-2}$ at T = 300 K, respectively[16]. Figure 2b also shows that $β$ increases linearly with # of layers, up to 5 layers, with a slope of 0.017 nN-nm$^2$/layer, close to the measured value of monolayer graphene, $β$ = 0.0199 nN-nm$^2$. This increase with layer number suggests that the strength of the force is increasing in an integer manner as additional graphene layers are added. This is consistent with the additive nature of the vdW force[1, 30]. Our results are interesting in the context of recent experiments where an AFM tip was pulled off of a graphene substrate where the pull-off force was observed to depend on the number of graphene layers in suspended membranes[34], but not on graphene supported by a substrate[34-36]. Despite this similarity in response, we note that pull-off experiments are well-known to be different mechanistically than the pull-in experiments of our study.

In addition to vdW force, the interfacial forces can be from capillary or electrostatic forces. The capillary forces take effect when graphene membranes or the substrate are covered with liquid films and the liquid films touch, and the force can be described by $P_{att} \propto 1/h$ [1, 30, 31]. However, we assume that the capillary force is not a likely candidate for the interfacial forces causing the pull-in phenomenon because absorbed liquid films of 10 nm thickness are unlikely to form between graphene membranes and the substrate[37, 38]. The electrostatic interaction, which can arise from image charges, work function differences or patch potentials can be described by $P_{att} \propto 1/h^2$ [1, 39, 40]. To further study the power law model considering different origins of the interaction, we varied the geometry of the annular ring. The pull-in distance for 1-4 layers graphene membranes with an identical outer diameter but a different inner diameter is shown in Fig. 3. The pull in distance shows a slight increase with increasing $b$. A theoretical calculation based on our analytical model using $P_{att} = \beta/h^4$ and the calculated values of $\beta$ in Fig.2b, is shown as a black shaded line in Fig. 3. The boundaries of the shaded lines show the range of values for $S_0 = 0.03 – 0.09$ N/m[29].

To determine if electrostatic forces play a significant role in our measurements, we fit the data in Fig. 1e and Fig. 3 with a model in which an electrostatic force takes the form, $P_{att} = \alpha/h^2$, and we use the same strategy to determine $\alpha$ as was used to calculate $\beta$ above. Doing so for the monolayer devices in Figure 1e, gives $\alpha = 0.49$ pN (for $S_0 = 0.07$ N/m). We can also use these values of $\alpha$ to fit the data in Fig. 3. This is shown as a shaded blue line which fits poorly to the data. A good fit would require that $\alpha$ increase with inner post diameter for all the devices measured, while no such assumption is needed for $\beta$. To fit all of our measured pull-in distances (51 devices in 17 geometries from 5 different chips) using an electrostratic force model requires that $\alpha$ values vary from $0.15 – 1.79$ pN across all the devices. A model based on an inverse 3$^{rd}$ power dependence was also examined and does not fit all the data as well as the inverse fourth power dependence (see Supplementary Information).

To test the material dependence of the interfacial interaction with graphene, we also carried out experiments where we measured the pull-in distance between graphene and a gold coated annular ring that were electrically contacted and grounded (see supplementary information). 2-5 layers graphene membranes (17 devices in 6 similar geometries from 4 chips) were measured. The pull in distance varied between 9 nm and 18 nm for annular rings with a = 1-1.75 µm and b = 0.15-0.6 µm, slightly larger than the measured pull-in distances for uncoated $SiO_x$ posts of a similar geometry. Using the same theoretical analysis as with the graphene/$SiO_x$ data, we determined the average value and standard deviation of $\beta$ / # of graphene layers between the Au coated post and electrically grounded graphene to be = 0.104 $\pm$ 0.031 nN-nm$^2$ / layer; these are about an order of magnitude higher than those for graphene interacting with SiO$_x$ (Fig. 4). The graphene/Au values agrees reasonably well with the theoretical predictions based on a Lifshitz formula of graphene interacting with gold at 15 nm separation, $\beta$ = 0.08 nN-nm.

In conclusion, we observed the pull in instability at 10nm-20nm distance on graphene by the attractive interfacial forces between graphene and SiO$_x$/Au, and found them to agree very well with a form $P_{att} = \beta/h^4$, consistent with recently calculated values of long range vdW forces between graphene and SiO$_x$ and graphene and gold. Furthermore, the strength of the force scales linearly with layer numbers, which is compatible with the additive nature of vdW forces. It is noteworthy that our experimental configuration is essentially a realization of a parallel plate geometry by self-alignment to measure interfacial forces acting on atomically thin, two-dimensional materials[41]. These experiments which provide a measurement of the magnitude and power law dependence of the interfacial forces at 10-20 nm separations between graphene and 2 common substrates can guide the development of nanomechanical devices from single and few layer graphene sheets where these forces are critical to their effective operation[6, 42, 43].

**Acknowledgements:**

We thank Rishi Raj for use of the Raman microscope. This work was supported by NSF Grants #0900832(CMMI: Graphene Nanomechanics: The Role of van der Waals Forces), #1054406(CMMI: CAREER: Atomic Scale Defect Engineering in Graphene Membranes), the DARPA Center on Nanoscale Science and Technology for Integrated Micro/Nano-Electromechanical Transducers (iMINT), the National Science Foundation

(NSF) Industry/University Cooperative Research Center for Membrane Science, Engineering and Technology (MAST), and in part by the NNIN and the National Science Foundation under Grant No. ECS-0335765.

**Supporting Information Available:** Fabrication Processes, Counting the Number of Graphene Layers, Analytical Model, Finite Element Simulations, Calculation of $β$, Calculation of $α$, $γ$, and Deformation of Graphene Membrane by vdw Force. This material is available free of charge via the Internet at http://pubs.acs.org.

## Figure Captions

**Figure 1: Measurement of the Pull-in Distance**

(a) (upper) Optical image of suspended a few layer graphene membrane in an annular ring geometry. (lower) Side view schematic of the suspended graphene on the annular ring.

(b) (upper) A 3d rendering of an AFM image of a pressurized graphene membrane in the annular ring geometry before delamination from the inner post. (lower) Side view schematic of the pressurized suspended graphene on the annular ring.

(c) (upper) A 3d rendering of an AFM image of a pressurized graphene membrane in the annular ring geometry after delamination from the inner post. (lower) Side view schematic of the pressurized suspended graphene delaminated from the inner post.

(d) A series of AFM line cuts through the center of a pressurized graphene membrane during pull in. The outer diameter, $2a = 3$ µm, and inner diameter, $2b = 0.5$ µm.

(e) Pull in distance, $h_0$, vs. number of layers for graphene membranes in an annular ring geometry with $2a = 3$ µm and $2b = 0.5$ µm. (upper left inset) Side view schematic of the graphene membrane right before and after pull in.

**Figure 2: Scaling of $\beta$ with Number of Layers**

(a) Center deflection, $h$, vs. pressure difference, $\Delta p$, calculated for a monolayer graphene membrane in the annular ring geometry with an outer diameter, $2a = 3$ µm, and inner diameter, $2b = 0.5$ µm. The red dashed line at $\Delta p = 1.68$ kPa corresponds to pull-in and the deflection at this point is $h_0 = 9.2$ nm. The black line corresponds to the analytical model and the blue line is a finite element analysis model.

(b) The calculated values of $\beta$ vs. number of layers using the data in (a) assuming a model where the force responsible for pull-in has the form $P_{att} = \beta/h^4$. The initial tension $S_0$ is assumed to be 0.07 N/m. A best fit line through the data is also shown which has a slope of 0.017 nN-nm$^2$/# of layer.

**Figure 3: Scaling of the Pull in Distance with $P_{att}$**

Pull in distance, $h_0$, vs. inner diameter, $2b$, for a) 1 layer b) 2 layer c) 3 layer d) 4 layer graphene flakes (verified by Raman spectroscopy) with identical outer diameter but different inner diameters. The black and blue shaded lines are the calculated results for 2 different power law dependences $P_{att} = \beta/h^4$ (black) and $P_{att} = \alpha/h^2$ (blue) with $S_0 = 0.03 – 0.09$ N/m. The values of $\beta$ and $\alpha$ are listed in supplementary material. a) (inset) Optical image of 2 of the measured monolayer devices. The scale bar = 5 µm.

**Figure 4: Modeled vdW force vs. Number of Layers for SiO$_x$ and Gold**

Measured $\beta$ / *Number* of graphene layers between SiO$_x$ and 1 layer graphene (solid red squares), 2 layer graphene (solid green circles), 3 layer graphene (solid blue up triangles), 4 layer graphene (solid cyan down triangles), 5 layer graphene (solid magenta diamond), and $\beta$ / *number* of graphene layers between Au and 2 layer graphene (hollow green circles), 3 layer graphene (hollow blue up triangles), 4 layer graphene (hollow cyan down triangles), and 5 layer graphene (hollow magenta diamond). The average and standard deviation of $\beta$ / *Number* of graphene layers between SiO$_x$ and graphene are $0.0179 \pm 0.0037$ nN-nm$^2$ / layer.

The average and standard deviation of *β / Number* of graphene layers between Au and graphene are $0.104 \pm 0.031$ nN-nm$^2$ / layer. Each data point corresponds to a separate device. (top left inset) Side view schematic of the pressurized suspended graphene on the annular ring with SiO$_x$ surface. (top right inset) Side view schematic of the pressurized suspended graphene on an Au coated annular ring.

**Figures**

Fig.1

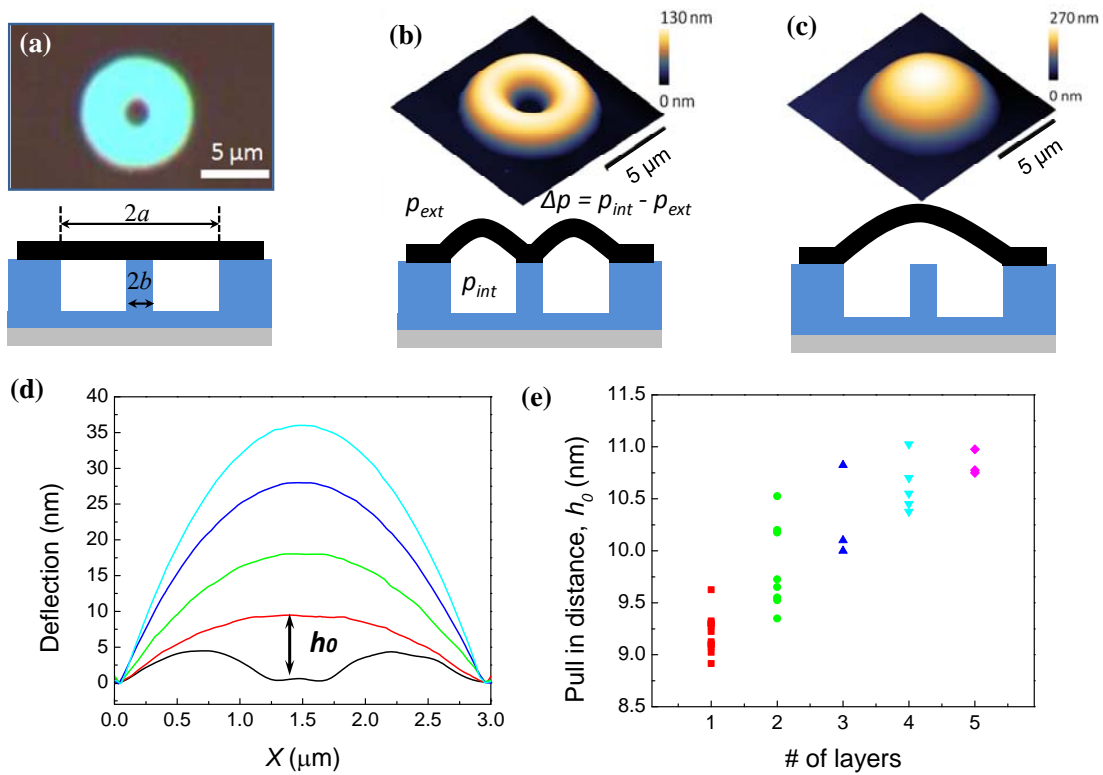

Fig.2

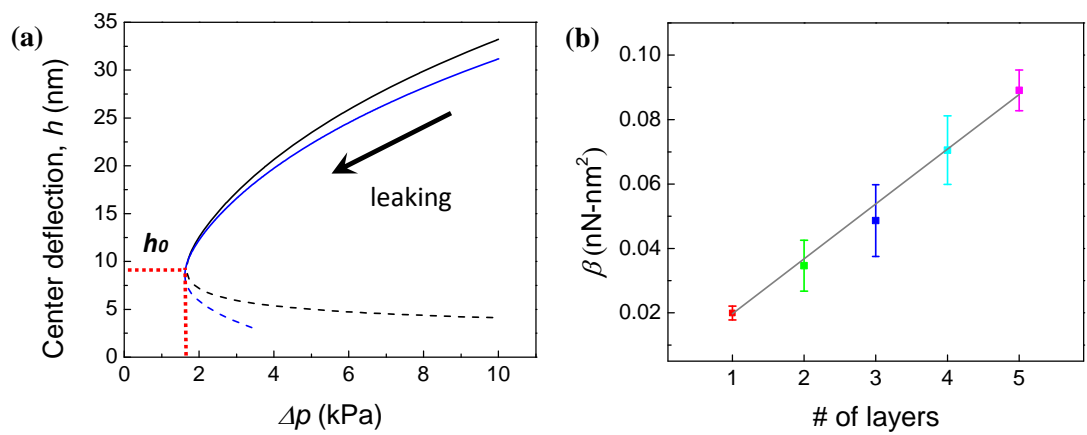

Fig.3

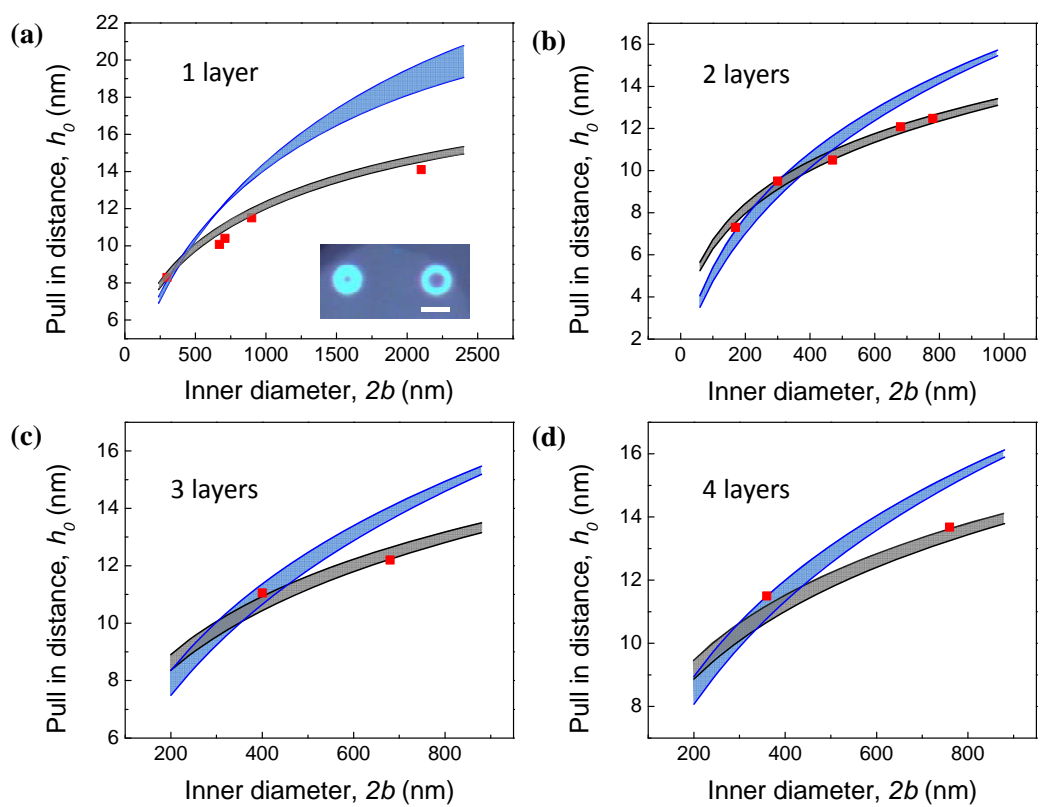

Fig.4

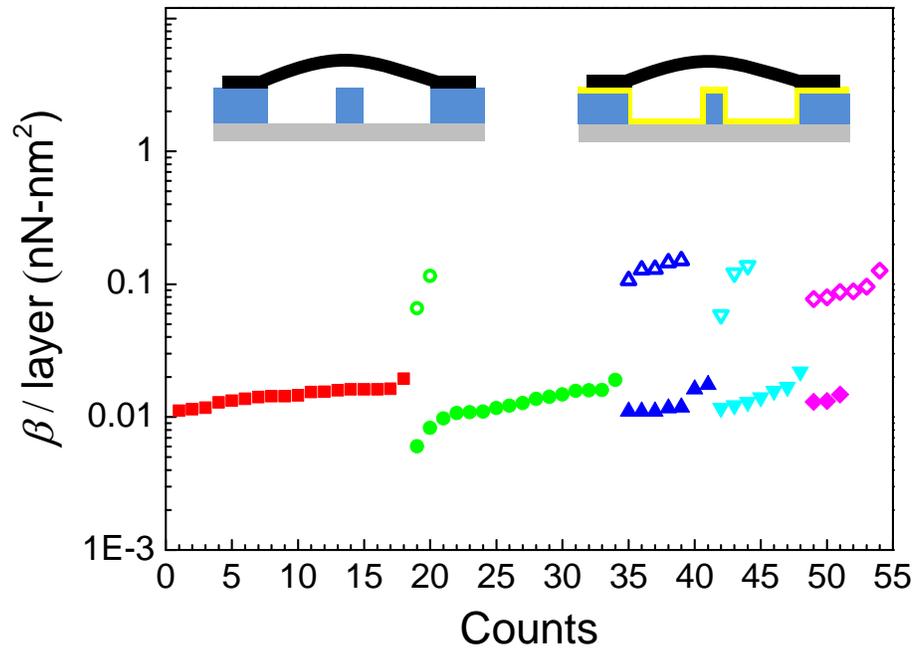

TOC Graphic

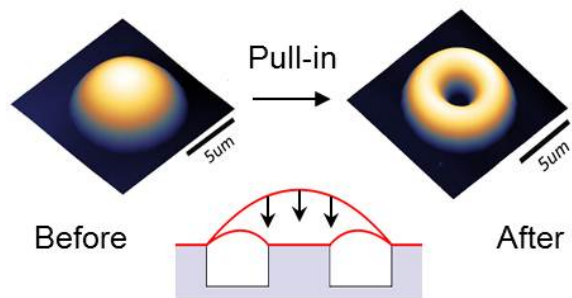

**Supplementary Information**

# Observation of Pull-in Instability in Graphene Membranes under Interfacial Forces

*Xinghui Liu[1], Narasimha G. Boddeti[1], Mariah R. Szpunar[2], Luda Wang[1], Miguel A. Rodriguez[3], Rong Long[1,4], Jianliang Xiao[1], Martin L. Dunn[5], and J. Scott Bunch[1]\**

[1]Department of Mechanical Engineering, University of Colorado, Boulder, CO 80309

USA

[1]Department of Mechanical Engineering, University of Colorado, Boulder, CO 80309

USA

[2]Department of Mechanical Engineering, University of Miami, Coral Gables, FL 33124

USA

[3]Department of Mechanical Engineering, Columbia University, New York, NY 10027

[4]Department of Mechanical Engineering, University of Alberta, Edmonton, Alberta T6G

2G8, Canada

[5]Singapore University of Technology and Design, Singapore, 138682

*email: jbunch@colorado.edu

## Fabrication Processes

Suspended graphene membranes were fabricated by a combination of standard photolithography, reactive ion etching and mechanical exfoliation of graphene. An array of annular cavities with designed dimensions was first defined by photolithography on an oxidized silicon wafer with a silicon oxide thickness of 90/285 nm. Reactive ion etching was then used to etch the annular rings into microcavities with a depth of 100-120 nm. After removal of photoresist with acetone and isopropanol, the chips were further cleaned in a Nanostrip bath at 60°C for 20 minutes. Thermal evaporation is used to deposit a layer of Cr/Au 5/10 nm for the Au coated annular rings. During the evaporation process, the chips are tilted at a 10~15° angle, so that the Cr/Au atoms can be deposited into the annular rings and cover the side walls. The large aspect ratio between the width and depth of the annular ring allows for a conformal metal deposition such that the post and the substrate are electrically contacted and grounded. Mechanical exfoliation of natural graphite using Scotch tape was then used to deposit suspended graphene sheets over the microcavities.

The pull-in distances in Fig. 1e were measured from two graphene flakes about 100 μm apart from each other on the same chip (Fig. S1). In the two graphene flakes, there were 13 one-layer, 9 two-layer, 5 three-layer, 5 four-layer, and 3 five-layer suspended membranes. For both the graphene/SiO$_x$ and the graphene/Au annular rings, the number of graphene layers was verified using Raman spectroscopy and optical contrast.

## Counting the Number of Graphene Layers

In order to count the number of graphene layers used in this study, we used optical contrast verified by Raman spectroscopy. Figure S1 (a) shows a graphene flake used in this study. The devices in Figure S1 (a) correspond to the devices in Figure 1e. The corresponding spots where Raman spectrum was taken for each device are shown as colored circles; red is 1 layer, green is 2 layers, blue is 3 layers, cyan is 4 layers and magenta is 5 layers. Figures S1 (b) shows the Raman spectrum taken from the spots of corresponding color in S1 (a), respectively. To verify the number of layers we found the ratio of the integrated intensity of the first order optical phonon peak and the graphene G peak (Fig. S1 (c))[1].

To measure the Raman spectrum on the gold coated samples, we patterned areas that contained no Au/Cr over which Raman spectrum of the graphene was taken without interference from the gold film. We patterned 5 μm circular discs between the annular wells using photolithography which masked the subsequent thermal evaporation of Au/Cr onto the $SiO_x$. After evaporation and lift-off, the protected areas contained no Au/Cr while all other areas of the wafer were covered with the Au/Cr film. We then used mechanical exfoliation to deposit the graphene and took the Raman spectrum of graphene through the 5 μm circular wells similarly to Fig S1. Figure S2 (a) shows a few layer graphene flake on Au/Cr coated wafer. The larger circles are locations where there is no Au/Cr and only $SiO_x$ with or without graphene. The blue circle corresponds to the location where Raman spectrum was taken. The number of graphene layers is verified using the same method as previously introduced.

## Analytical Model

We developed a simple analytical model based on membrane mechanics to describe the interrelationship of the system parameters in the experiment and we use it inversely with the measurements to infer the operant surface forces[2]. The symbols used in our approach are:

$b$ = Post Radius

$a$ = Outer Radius of the cavity

$E$ = Young's Modulus

$t$ = thickness

$v$ = Poisson's Ratio

$S$ = Total Tension/Membrane Force in the radial direction

$S_r$ = Incremental tension in the radial direction

$S_t$ = Incremental tension in the tangential direction

$S_0$ = Initial equi-biaxial tension

$\Delta P$ = Pressure exerted by the difference of gas pressures inside and outside the cavity

$P_{att}$ = Pressure due to the post-graphene interactions

$r$ = Radial Co-ordinate, $0 < r < a$

$w$ = Deflection of the membrane, as a function of $r$

$h$ = Deflection at $r = 0$

The key assumptions of our treatment are:

1) The membrane tension $S$ is uniform.

2) The pressure due to the surface forces acting between the post and the membrane, $P_{att}$, is uniform. This is reasonable if the membrane curvature is small. This is the case when the post is small compared to the overall size of the cavity.

In order to understand the validity and impact of these assumptions, we also carry out high-fidelity finite element (FE) simulations where they are removed; these are described in the next section.

Force equilibrium in the vertical direction gives (see Figure S3 (a)):

$$(\Delta P - P_{att})r^2 = -2\,S\,r\frac{dw}{dr} \quad r < b$$

$$\Delta P\,r^2 - P_{att}b^2 = -2\,S\,r\frac{dw}{dr} \quad r \geq b$$

$$S = S_r + S_0$$

The negative sign on the right hand side is due to $dw/dr$ being negative. Integrating with respect to $r$ with appropriate limits, yields:

$$w = h - \frac{\Delta P - P_{att}}{4S}r^2 \quad r < b$$

$$w = w(r = b) + \frac{1}{4S}\left(P_{att}b^2 \log\left(\frac{r^2}{b^2}\right) - \Delta P(r^2 - b^2)\right) \quad r \geq b$$

Due to continuity of $w$ at $r = b$ we obtain:

$$w = h - \frac{\Delta P - P_{att}}{4S} r^2 \quad r < b$$

$$w = h + \frac{1}{4S}\left(P_{att} b^2 \log\left(\frac{r^2}{b^2}\right) + P_{att} b^2 - \Delta P \, r^2\right) \quad r \geq b$$

Applying the boundary condition $w(r = a) = 0$, yields:

$$h = \frac{1}{4S}\left(\Delta P \, a^2 - P_{att} b^2 \left(1 + \log\left(\frac{a^2}{b^2}\right)\right)\right) \quad (1)$$

Finally,

$$w = \frac{1}{4S}\left(\Delta P(a^2 - r^2) - P_{int}(b^2 - r^2) - P_{att} b^2 \log\left(\frac{a^2}{b^2}\right)\right) \quad r < b \quad (2)$$

$$w = \frac{1}{4S}\left(\Delta P(a^2 - r^2) + P_{att} b^2 \log\left(\frac{r^2}{a^2}\right)\right) \quad r \geq b \quad (3)$$

We assume that the membrane is in an equi-biaxial state, then $S_r = S_t$ and $\epsilon_r = \epsilon_t = \frac{S}{Et/(1-v)}$ and:

$$\epsilon_r + \epsilon_t = \frac{du}{dr} + \frac{u}{r} + \frac{1}{2}\left(\frac{dw}{dr}\right)^2 = \frac{2 S_r}{Et/(1-v)}$$

Integrating with respect to an area element $2\pi r dr$ over $(0, a)$, yields:

$$\int_0^a d(ur) + \int_0^a \frac{r}{2}\left(\frac{dw}{dr}\right)^2 dr = \frac{2 S_r}{Et/(1-v)} \int_0^a r dr$$

The first integral on the LHS is zero due to the boundary conditions and thus:

$$S_r S^2 = \frac{Et}{32\,a^2(1-v)}\left((\Delta P - P_{att})^2 b^4 + \Delta P^2\left(a^4 - b^4\right) + P_{att}^2 b^4 \log\left(\frac{a^4}{b^4}\right)\right.$$
$$\left. - 4\,\Delta P\,P_{att} b^2\left(a^2 - b^2\right)\right) \qquad (4)$$

In order to obtain the condition for pull-in we eliminate $S_r$ and $S$ from eqs. (1) and (4) results in an equation for $h$ in terms of $a$, $b$, $Et$, $v$, $\beta$, $S_0$, $P_{att}$ and $\Delta P$; in our experimental configuration all of these are known except $\Delta P$ and $S_0$. When we specify a particular value of $S_0$ this yields an expression for the load-deflection behaviour, i.e., $\Delta P$ vs. $h$.

$$\frac{Et}{32\,a^2(1-v)}\left((\Delta P - P_{att})^2 b^4 + \Delta P^2\left(a^4 - b^4\right) + P_{att}^2 b^4 \log\left(\frac{a^4}{b^4}\right)\right.$$
$$\left. - 4\,\Delta P\,P_{att} b^2\left(a^2 - b^2\right)\right)$$
$$+ \left(S_0\left(\frac{1}{4h}\left(\Delta P\,a^2 - P_{att} b^2\left(1 + \log\left(\frac{a^2}{b^2}\right)\right)\right)\right)^2\right)$$
$$= \left(\frac{1}{4h}\left(\Delta P\,a^2 - P_{att} b^2\left(1 + \log\left(\frac{a^2}{b^2}\right)\right)\right)\right)^3 \qquad (5)$$

Consistent with the van der Waals (vdW) form, we assume $P_{att}$ is given by a power law of the form,

$$P_{att} = \frac{\beta}{h^4}$$

The pull-in condition occurs at the limit point:

$$\frac{d\Delta P}{dh} = 0 \qquad (6)$$

which yields a unique $\Delta P$ and $S_0$ when $\beta$ and $h$ are specified.

## Finite Element Simulations

To validate the analytical model, we also carried out high-fidelity finite element simulations of the experimental configuration using the code Abaqus where we remove the assumptions used to develop the analytical model. The model used in the simulations is shown in Figure S3 (b). Axisymmetric shell elements (that permit both bending and membrane behaviour) were used and the Young's modulus and Poisson's ratio were set to 1 TPa[3] and 0.16[4] respectively. The outer edge of the membrane is pinned and the substrate/post is modelled as a fixed analytical rigid body. Since it is known that pressurized graphene behaves like a membrane and bending plays a negligible role in its mechanics[8,9], the value of the bending modulus and slope near the boundary is found to be irrelevant in these simulations. A prescribed initial tension is applied and the attractive interactions between the substrate and the membrane are modelled as surface-to-surface contact/adhesive interactions with the substrate being the master surface. The contact interaction properties are supplied through the user subroutine "UINTER" of Abaqus[10]. The slave nodes experience a tensile (attractive) contact stress ($\sigma_z$) only in the vertical direction given by,

$$\sigma_z(r) = -\frac{\beta}{w(r)^4}$$

Here, $\beta$ is a parameter and $w$ is the deflection of the node measured from the substrate. Both $\sigma_z$ and $w$ are functions of the radial position, in contrast to the analytical model where they are assumed to be independent of position.

The simulation is split into two steps – both static steps with nonlinear geometric effects included. In the step 1, the contact/adhesive interactions are suppressed and the membrane is allowed to deform under the influence of a uniform pressure load acting on the entire area of the suspended membrane. The magnitude of this load is set such that the deflection is just high enough to neglect the interaction pressure if the interactions were not suppressed. This simulates the state of affairs at the beginning of the experiment before the gas begins to leak from the cavity. In the second step, which is a Static-Riks step[10], a second uniform pressure load is added with the same magnitude as the previous pressure load but in the opposite direction and the surface interactions between the substrate and the membrane are switched on. Hence at a given increment during the step, apart from the force due to the contact interactions, the membrane has the uniform pressure load from the previous step and a uniform pressure in the opposite direction whose value is given by the load proportionality factor. The superposition of these two uniform pressure loads mimics the leaking of the gas in the experiment. As the simulation progresses, the load across the membrane decreases and it comes closer to the substrate. This increases the interaction between the post and membrane. The results of this step are plotted in Figure 2a of the main text. It can be seen that the load across the membrane initially decreases until a limit point is reached and then it starts increasing. The limit point gives the pull-in distance and the pressure at which it occurs. The configurations below the limit point can't be achieved in a load controlled experiment, but suggest that system has two possible equilibrium configurations at a given pressure load greater than the pull-in pressure. Careful comparison of the analytical and finite element simulation results (Fig. S4) shows that the analytical result is an accurate description of the physical phenomena as long as the substrate/post size is small compared to the size of the suspended membrane.

## Calculation of $\beta$

Using the analytical model described above, we calculate the values of $\beta$ assuming a range of initial tension, $S_0$. Previous results on mechanically exfoliated monolayer and few layer graphene found $S_0$ in the range of 0.03 - 0.15 N/m where the average values was $S_0$ =0.07 N/m[5–7]. Figure S5a shows calculated $\beta$ for different $S_0$ (0.03, 0.05, 0.09 N/m). This range also marks the shaded boundaries for the theoretically calculated pull-in distance in Fig. 3.

## Calculation of $\alpha$, $\gamma$

The same analytical model used to calculate $\beta$ can be applied to $\alpha$ and $\gamma$, where $\gamma$ is a constant similar to $\alpha$ and $\beta$ assuming $P_{att} = \gamma/h^3$. The inverse cubic dependence for the interfacial interactions can arise due to vdW interactions between thick graphene membranes and the substrate. Calculated $\alpha$ and $\gamma$ with $S_0$ = 0.03, 0.05, 0.09 N/m is shown in Fig. S5b and Fig. S5c. The calculated $\alpha$ for all the devices measured is plotted in Fig. S6a assuming $S_0$ = 0.07 N/m. The same analysis is done with $\gamma$ shown in Fig. S6b. We also plot pull-in distance ($h$) versus post diameter ($2b$) for this power law and compare it with $P_{att} = \beta/h^4$ and the experimental data in Fig. S7. Even though the plot fits experimental data closely for 2-4 layers graphene membrane, it does not fit the data from monolayer graphene membranes as well.

### Deformation of graphene membrane by vdw force

The extreme flexibility of the suspended graphene coupled with the large magnitude of the interfacial force at these short separations shows up as a statically deformed membrane right before pull-in for some devices. This is especially evident for a graphene membrane with a small inner post – more localized force- and a large outer diameter – more flexible graphene (Fig. S8). The AFM image shows a graphene membrane locally deformed at its center shortly before pull-in (Fig. S8a). The AFM line cut through the center (Fig. S8b) shows this deformation to be about 2 nm. This deformation is further verified by the analytical model which shows a number of stable configurations for graphene membranes deformed by $P_{att}$ at these dimensions and separations (Fig. S8c).

**Supplementary References:**

## Supplementary Figure Captions

**Figure S1: Determining the number of layers**

(a) Optical image showing one of the graphene flakes corresponding to some of the samples measured in Fig. 1e. The colored circles denote the location at which Raman spectroscopy was taken (black-1 layer, red-2 layers, green-3 layers, blue-4 layers, and cyan-5 layers).

(b) Raman spectrum for the graphene flake in (a). The color of each curve corresponds to Raman spectrum taken at the corresponding colored circle in the optical image.

(c) Ratio of the integrated intensity of the first order silicon peak I(Si) and the graphene G peak, I(G) (i.e. I(G)/I(Si) for the Raman spectrum in (b).

**Figure S2: Additional Raman spectrum**

(a) Optical image showing a few layer graphene flake on Au coating. The blue circles denote the location at which Raman spectroscopy was taken.

(b) Raman spectrum for 2-5 layers graphene flakes on Au coating through 5 μm wells.

(c) Ratio of the integrated intensity of the first order silicon peak I(Si) and the graphene G peak, I(G) (i.e. I(G)/I(Si) for the Raman spectrum in (b).

**Figure S3: Schematic of the model**

(a) Schematics showing the equilibrium condition for the two regions of the membrane.

(b) Schematic of the model used for finite element analysis simulations.

**Figure S4: Comparison of analytical solution and finite element simulations**

(a) (a) Plots comparing p vs h behavior as obtained from the FE simulations (solid curve) and the analytical calculations (dashed curve) with a = 1.5 μm, b = 0.25 μm, Et = 340 N/m, ν = 0.16, S0 = 0.07 N/m and β = 0.02 nN-nm2,

(b) The deflection profiles at different pressures (solid – FE, dashed – Analytical) (Red – 10.38 KPa, Blue – 6.12 KPa, Green – 1.72 KPa and Magenta – 2.61 KPa). For convenience, the corresponding points on p vs h plot are also shown. (c) and (d) The same as (a) and (b) except b = 0.75 μm. The different pressures used in this case are: Red – 10.39 KPa, Blue – 6.14 KPa, Green – 2.63 KPa and Magenta – 3.70 KPa.

**Figure S5: $\alpha$, $\beta$, $\gamma$ vs. number of layers**

(a) The calculated values of $\beta$ vs. number of layers assuming a model where the force responsible for pull-in has the form $P_{att} = \beta/h^4$ with different initial tension $S_0 =$ 0.03 N/m, $S_0 = 0.05$ N/m, $S_0 = 0.09$ N/m

(b) The calculated values of $\alpha$ vs. number of layers assuming a model where the force responsible for pull-in has the form $P_{att} = \alpha/h^2$ with different initial tension $S_0 =$ 0.03 N/m, $S_0 = 0.05$ N/m, $S_0 = 0.09$ N/m

(c) The calculated values of $\gamma$ vs. number of layers assuming a model where the force responsible for pull-in has the form $P_{att} = \gamma/h^3$ with different initial tension $S_0 =$ 0.03 N/m, $S_0 = 0.05$ N/m, $S_0 = 0.09$ N/m

**Figure S6: α, γ for all devices measured.**

- **(a)** Calculated $\alpha$ for all the devices measured assuming $S_0 = 0.07$ N/m; (same color scheme as Fig. S5a).

- **(b)** Calculated $\gamma$ for all the devices measured assuming $S_0 = 0.07$ N/m; (same color scheme as Fig. S5a).

**Figure S7: Scaling of the Pull in Distance with $P_{att}$**

Pull in distance, $h_0$, vs. inner diameter, $2b$, for a) 1 layer b) 2 layer c) 3 layer d) 4 layer graphene flakes (verified by Raman spectroscopy) with identical outer diameter but different inner diameters. The black and blue shaded lines are the calculated results for 2 different power law dependences $P_{att} = \beta/h^4$ (black) and $P_{att} = \alpha/h^3$ (blue) with $S_0 = 0.03 - 0.09$ N/m. The values of $\beta$ and $\gamma$ are listed in Fig. S5. a) (inset) Optical image of 2 of the measured monolayer devices. The scale bar = 5 μm.

**Figure S8: Deforming a Graphene Membrane with the vdw Force**

- **(a)** An atomic force microscope image showing a close up view of the top part of the pressurized graphene membrane right before pull-in showing the deformation at the center of the membrane resulting from the vdw force.

- **(b)** A line cut through the center of the image in (a).

- **(c)** Calculated deflection vs. position through the center of a graphene membrane using the analytical model, for varying $S_0$.

# Figure

Fig.S1

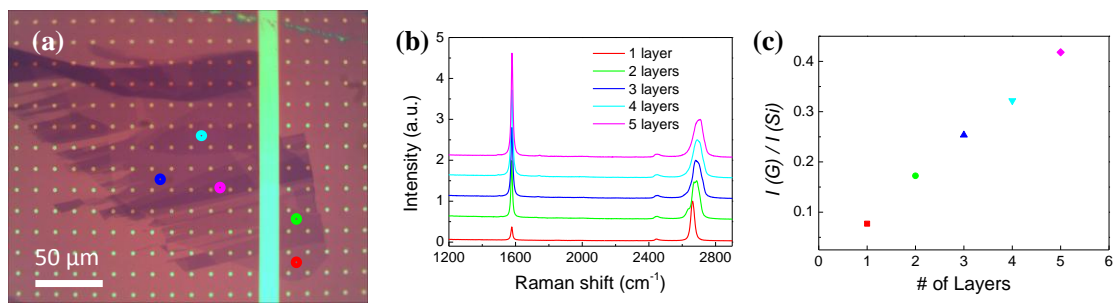

Fig.S2

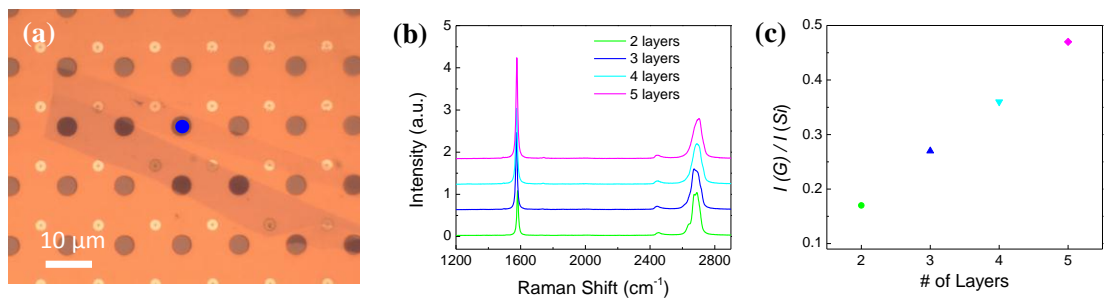

Fig.S3

**(a)**

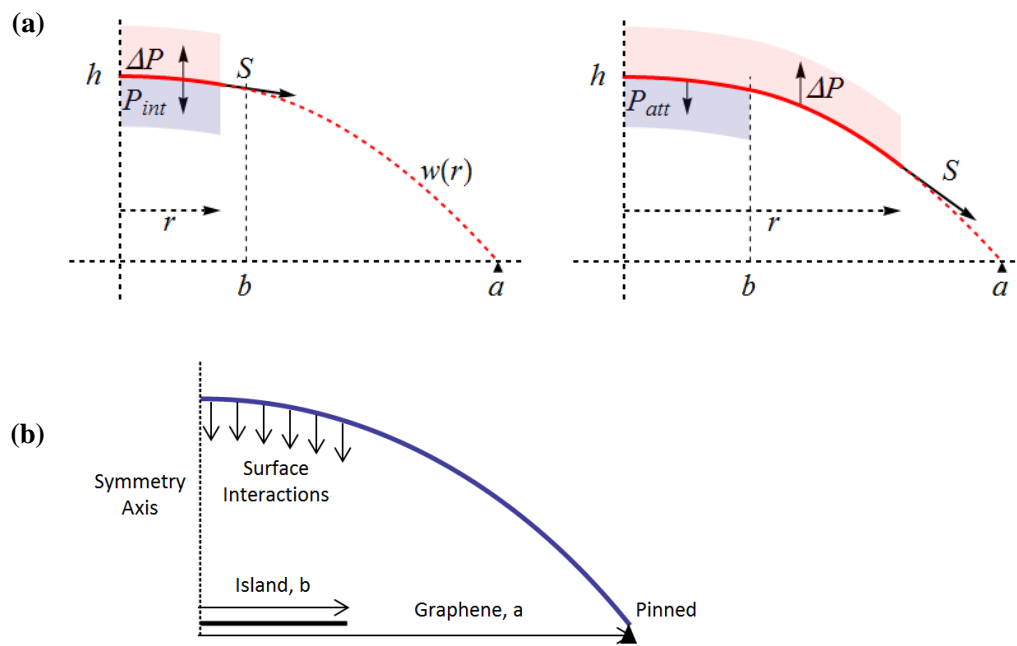

**(b)**

Fig.S4

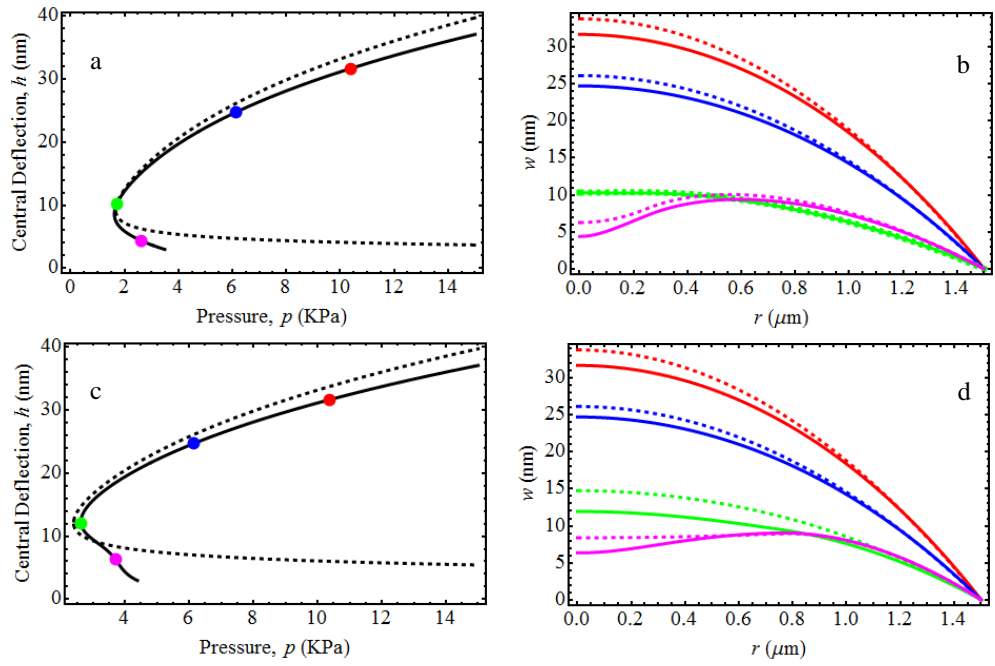

Fig.S5

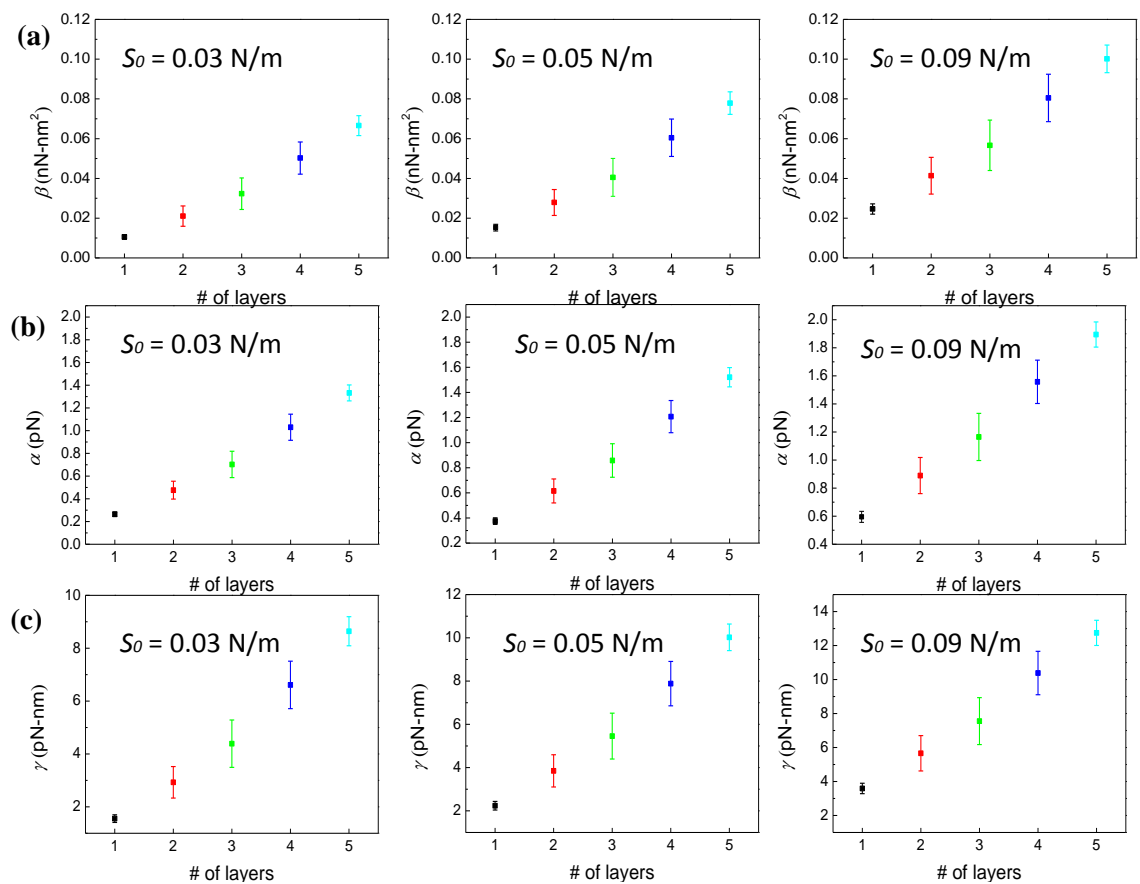

Fig.S6

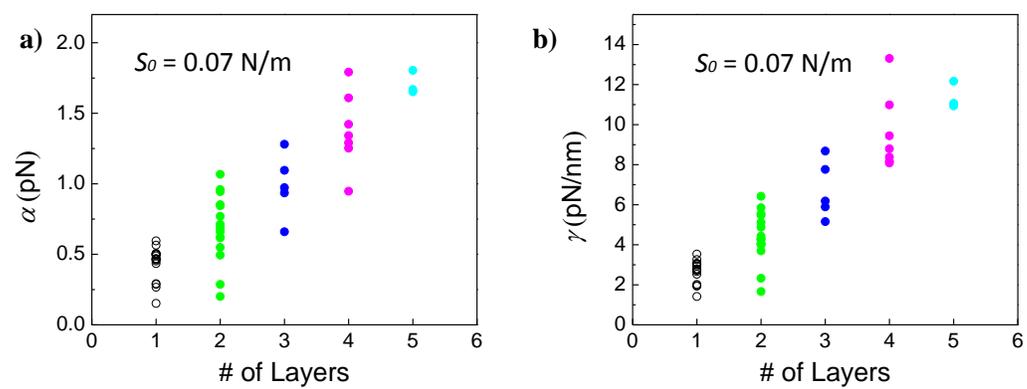

Fig.S7

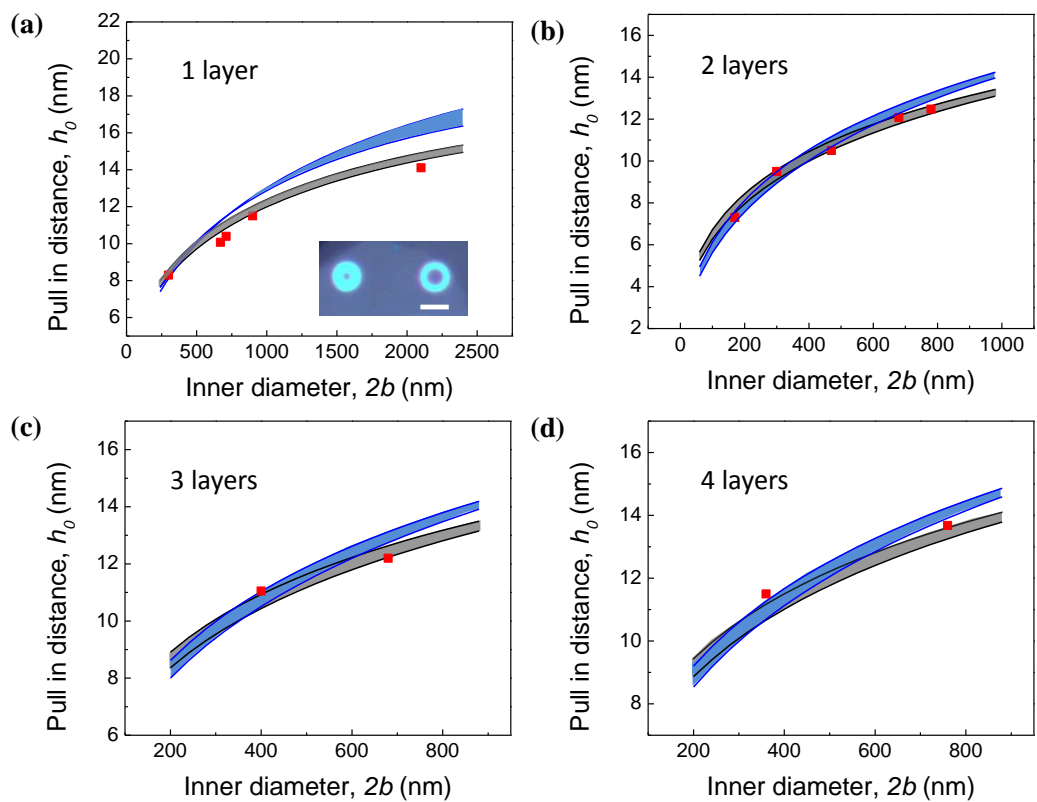

Fig. S8

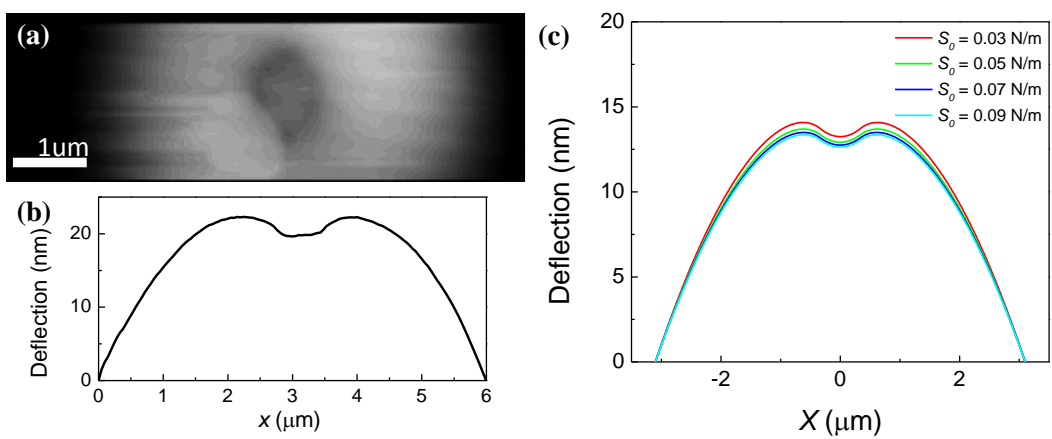